\documentstyle[prl,aps]{revtex}
\twocolumn
\begin{document}
\draft
\title{Entropic Origin of the Growth of Relaxation
Times in Simple Glassy Liquids}
\author{Chandan Dasgupta\cite{jnc}}
\address{Department of Physics, Indian Institute of Science, Bangalore 
560012, India}
\author{Oriol T. Valls}
\address{School of Physics and Astronomy and Minnesota 
Supercomputer Institute, \\ University of Minnesota, 
Minnapolis, Minnesota 55455}
\date{\today}
\maketitle
\begin{abstract}
Transitions between ``glassy'' local minima of a model free-energy functional
for a dense hard-sphere system are studied numerically 
using a ``microcanonical''
Monte Carlo method that enables us to obtain 
the transition probability as a function of
the free energy and the Monte Carlo ``time''. The growth of the
height of the effective free energy barrier with 
density is found to be consistent with
a Vogel-Fulcher law. The dependence of the transition 
probability on time indicates
that this growth is primarily due to entropic effects
arising from the difficulty of finding low-free-energy saddle points
connecting glassy minima.
\end{abstract}

% body of paper here

The dynamic behavior of supercooled liquids near the glass transition 
\cite{rev} is one of the
most enigmatic problems of condensed matter physics. 
The most dramatic feature of the dynamics near the glass 
transition in so-called ``fragile'' systems \cite{angell} is 
an extremely rapid growth of the relaxation time 
$\tau$ which is reasonably well-described by the Vogel-Fulcher law \cite{vf},
$\tau \propto \exp[C/(T-T_0)]$, where $T_0 < T_g$, the conventionally defined
glass transition temperature at which the viscosity 
attains a value of 10$^{13}$ P. The apparent divergence of $\tau$ 
at $T_0$ has led to speculations about the possibility
of a true thermodynamic transition at this temperature. This 
is also suggested by the observation 
\cite{kauz} that the temperature $T_K$ (the so-called
Kauzmann temperature) at which the entropy difference between 
the supercooled liquid and the equilibrium crystalline solid 
extrapolates to zero is very close to $T_0$. The
closeness of $T_0$ and $T_K$ suggests that the growth 
of the relaxation time near the glass transition is primarily entropic 
in origin. Heuristic arguments that attempt to
relate the Vogel-Fulcher law to entropic effects have been proposed by several
authors \cite{gdm,ktw}. However, we are not aware of any 
calculation that provides an explicit demonstration of such 
effects in simple model
liquids. 

In this Letter, we describe the results of a numerical investigation 
that provides direct evidence for an entropic origin of the growth of 
the relaxation time in a dense
hard sphere fluid near the glass transition. Our computations are 
based on a model free energy functional \cite{ry} for the hard sphere system. 
We use a novel ``microcanonical''
Monte Carlo (MC) method to study transitions between different ``glassy'' 
minima of a discretized version of this free energy functional. 
We determine the probability of transition from one minimum to another as a 
function of the free energy
increment $\Delta f$ (the excess free energy per particle
measured from that at the original minimum) and MC 
``time'' $t$. This allows us to define an effective barrier height that
depends weakly on $t$. We find that the growth of this effective 
barrier height with increasing density is consistent with a Vogel-Fulcher 
form appropriate for a hard-sphere system. Our numerical results about how 
the dependence of the effective barrier height on $t$ changes as the 
density is increased indicate clearly that the growth of the barrier height 
(and the consequent growth of the relaxation time)
is primarily due to entropic effects arising from an increase in the 
difficulty of finding low free-energy paths 
(``saddle points'') that connect one glassy local minimum with
another.

The free energy functional used in our study is of
the form proposed by Ramakrishnan and
Yussouff \cite{ry}:
\begin{eqnarray}
F[\rho] &=& F_l[\rho_0]+ k_B T \left[ \int{d {\bf r}\{\rho({\bf r})
\ln (\rho({\bf r})/\rho_0)-\delta\rho({\bf r})\} } \right. \nonumber \\
&-& \left. (1/2)\int{d {\bf r} \int {d{\bf r}^\prime
C({|\bf r}-{\bf r^\prime|}) \delta \rho ({\bf r}) \delta
\rho({\bf r}^\prime)}} \right],
\label{ryfe}
\end{eqnarray}
where $\delta \rho ({\bf r})\equiv \rho({\bf r})-\rho_0$ is the deviation
of the time-averaged local
number density $\rho({\bf r})$ from its value $\rho_0$ in the
uniform liquid state, $F_l$ is the free energy of the uniform liquid, 
$T$ is the temperature, and $C(r)$ is the direct pair correlation 
function \cite{hm86} of the uniform liquid at density $\rho_0$.
$C(r)$ is expressed in terms of the dimensionless
density $n^*\equiv \rho_0 \sigma^3$ ($\sigma$ is the hard sphere
diameter) through the Percus-Yevick
approximation \cite{hm86} which is expected to be adequate if $\rho_0$ is
not very high.

The discretized version of this free 
energy functional exhibits\cite{cd92} a large number of ``glassy'' minima 
(local minima of $F$
at which the density is inhomogeneous but aperiodic) for $n^* > n^*_f$ where 
$n^*_f \simeq 0.85$
is the density at which equilibrium crystallization 
occurs. Numerical studies \cite{lvd93,dv94}
of Langevin equations appropriate for this system show that the dynamic
behavior is governed by thermally activated transitions among 
these glassy minima if $n^*$ exceeds a ``crossover'' value that is close 
to 0.96. The time scales for such transitions were estimated 
from a standard MC method in Ref.\cite{dv96} and found to 
rapidly increase with increasing density. Here, we have
used a different numerical method that is more efficient than the canonical MC
method and provides information about the origin of the growth of the 
time scale of these thermally activated transitions. The
increase of this time scale may be due to one or both of two factors:
(a) an increase of the height of the lowest
free-energy barrier that separates a glassy minimum from the others; and (b) an
increase of the difficulty in finding low-free-energy paths to other minima. 
Considering the free energy functional as an effective 
Hamiltonian for the system, 
these two factors may be called ``energetic'' and ``entropic'' \cite{footn1}, 
respectively.
The canonical MC method does not provide much information about the relative
importance of these two factors in the observed growth of the time scales. 
As described below, the numerical method used in this study 
allows us to distinguish between energetic and entropic effects. It also allows
us to follow the growth of the barrier-crossing time scale over 
about ten decades,
which would not be possible in a canonical MC calculation.

We discretize our system on a cubic lattice of size $L^3$ and mesh constant
$h$ with dimensionless density variables defined as
$\rho_i \equiv \rho({\bf r}_i)h^3$. Periodic boundary conditions are
used and the constraint that
the sum of the variables $\rho_i$ must be a constant $N$, the number of 
particles in the sample, is enforced during the simulation.
We define a dimensionless free energy 
per particle, $f[\rho]$, as:
\begin{equation}
f[\rho]= \beta F[\rho]/(n^* L^3 a^3)
\label{f}
\end{equation}
where $a$ is the ratio $h/\sigma$, and $\beta =1/k_BT$.

Our numerical method, which may be called ``microcanonical'' MC 
if the free energy
functional is considered to be an effective Hamiltonian,
involves the following steps. 
Each run is started from a glassy local minimum of the free energy.
We choose a trial value of what we call the free energy increment
$\Delta f$ and then
perform a MC simulation in which we
sweep the sites $i$ of the lattice sequentially. At each step and site,
we pick another site $j$ at random from 
the ones that lie within a distance $\sigma$
from the site $i$. We then attempt to change the values
of $\rho_i$ and $\rho_j$ to $p(\rho_i+\rho_j)$ and $(1-p)(\rho_i+\rho_j)$,
where $p$ is a random number distributed uniformly in $[0,1]$. 
The atempted change
is accepted {\em only if the dimensionless free energy
after the change is less than} $ F_{min} + N\Delta f$ where
$F_{min}$ is the dimensionless free energy $\beta F$ at
the minimum where the simulation is started.
This procedure generates a random sampling of configurations whose free energy
lies within $N \Delta f$ of that of the glassy minimum under consideration. 
The simulation proceeds up to a maximum ``time'' $t_{m}$, of MC steps per
site. At regular intervals along the evolution of the system, we 
use a minimization
procedure \cite{cd92} to determine whether the system has moved to 
the basin of 
attraction of a different local minimum of the free energy.
Obviously, if $N\Delta f$ is 
smaller than the lowest free-energy barrier between the starting minimum
and any other minimum, the system 
remains in the basin of attraction of the starting minimum. As
$\Delta f$ is increased, one begins to find transitions to 
other accessible minima, that is, minima
that the system can find within a time $t \leq t_m$, which
are separated from the initial one by a barrier of height 
less than $N\Delta f$.
Repeating this procedure a number of times (typically 10-20) for a fixed
set of values of $n^*$, $\Delta f$ and $t_m$, we obtain 
$P(n^*,\Delta f, t)$, the probability 
of a transition to a different minimum within time $t$ for
free energy increment $\Delta f$, as the fraction of the 
number of runs in which
a transition is found. This probability is calculated
for a suitable range of values of $n^*$, $\Delta f$, and $t$, and
the whole procedure is repeated for several glassy minima of 
the free energy (see below). 
We define a ``critical'' value, $\Delta f_c(n^*,t)$, of
the free energy increment as the value of $\Delta f$ for which 
$P(n^*,\Delta f,t) = 0.5$. 
Clearly, $N \Delta f_c$ represents an effective barrier height for
transitions to other local minima. 
This is the quantity that we
use to present our results.

We have used two sizes, $L$ = 15 and 
$L$ = 12. In the first case we have taken $a$ = 1/4.6 so that $L$ and
$a$ are incommensurate with a close-packed lattice and no crystalline minimum
of the free energy is found. The total number of inhomogeneous minima 
is then about 10 and all of them exhibit glassy structure as determined
by the two-point correlation function of the local density. 
The minima we have used as our starting point in this case
were also used in Ref.\cite{dv96}. These are the minima 
to which the system moves \cite{dv94}
in the course of its time evolution under Langevin
dynamics \cite{lvd93} when it is started from the uniform liquid state.
For $L$ = 12 we took $a$ = 0.25 so that the sample is commensurate. 
It admits a crystalline minimum that has the
lowest free energy for the values of $n^*$ considered here. The
number of glassy minima is substantially larger (about 30) in this sample. 
Out of those we chose a few with structure similar to
that of the minima of the $L$ = 15 sample.
For both cases, the minima found at lower densities 
were ``followed'' to higher densities
by running the minimization program at the higher density using the
lower density configuration (which is of course not a minimum at the
higher density) as the starting point. The values of $t_m$ 
are 15,000 for $L$ = 15 and 8,000 for $L$ = 12. The transition probability was
calculated at time intervals of 5,000 in the first case, 
and 2,000 in the second case.
In both cases, the density range covered was $0.94 \le n^* \le 1.06$.
Higher values of $n^*$ were not considered because the 
Percus-Yevick approximation then becomes
\cite{hm86} inaccurate.

Typical results for $P$ are shown in Fig.1 where data
for $L$ = 12 and $n^*$ = 1.04 are plotted for four different values of $t$.
The value of $\Delta f$ was incremented in steps of 0.05, which 
is also the estimated
uncertainty in the determination of $\Delta f_c$. 
The transition probability grows from zero as $\Delta f$ is increased,
and eventually saturates at one for sufficiently large values of $\Delta f$. 
For a fixed value of $\Delta f$, the transition probability increases as $t$ is
increased: transitions to other minima are more likely 
if the system is allowed to explore a larger number of configurations. 
Since $P$ is an increasing function of both $\Delta f$ and $t$,
$\Delta f_c(n^*,t)$  (the value of $\Delta f$ where $P$=0.5, as 
defined above) decreases as $t$ is increased. In agreement with
the previously observed \cite{dv96}
growth of the barrier-crossing time scale with $n^*$, we find that 
$\Delta f_c$ is an increasing function of $n^*$. 

The conclusion that entropic effects play a crucial 
role in the growth of the effective height of the
free energy barriers stems from the observation that
the $t$-dependence of $\Delta f_c$ becomes {\em stronger} as $n^*$
is increased (see Fig.2). 
It is intuitively obvious that the $t$-dependence of $\Delta f_c$ is
closely related to the probability of finding a path 
(``saddle point'') that connects
the starting minimum to a different one. 
If such paths were relatively easy to find, then the transition 
probability would be insensitive to the value of $t$ as long as
it is not very short. If, however, paths to other minima are few,
a large number of configurations have to be explored 
before one of them is found. The $t$-dependence 
of $\Delta f_c$ would then be more pronounced
and extend to larger values of $t$. To make the idea more concrete, 
we ignore the correlations (which are short-range in time) among 
the configurations generated in a
MC run and assume that they represent $t$ independent 
samplings of configurations with
free energy less than $F_{min}+N\Delta f$.  
Let us also assume that
the system does not return to the basin of attraction of 
the starting minimum after 
a transition to a different basin of attraction. We find
that a return to the original basin of attraction is indeed 
very rare. The transition probability may then be estimated
as $P(n^*,\Delta f,t) = 1- [1-p(n^*,\Delta f)]^{t} \simeq 1 - \exp(-tp)$, where
$p(n^*, \Delta f) \ll 1$ is the probability that a randomly 
chosen configuration with
$\beta F \le F_{min} + N \Delta f$ belongs in the basin of attraction of
a different minimum. One expects $p$ to be zero if $\Delta f \le
\Delta f_0(n^*)$ where $N \Delta f_0$ is the height of the lowest 
free energy barrier, and
$p = g(n^*,\Delta f - \Delta f_0)$ for $\Delta f > \Delta f_0$ where 
$g(n^*,x)$ grows 
continuously from zero as $x$ is increased from zero.
Combining this with the definition of $\Delta f_c$, we obtain the relation 
$g(n^*,\Delta f_c(n^*,t)-\Delta f_0(n^*)) = \ln 2/t$. 
Our observation that the difference
$\Delta f_c(n^*,t_1) - \Delta f_c(n^*, t_2)$ for fixed $t_1 < t_2$ 
{\em increases} with $n^*$ then leads to the conclusion that 
the function $g(n^*,x)$
{\em decreases} (i.e. the difficulty of finding 
paths to other minima increases) as
$n^*$ is increased at fixed  $x$.

The observed $t$-dependence of $\Delta f_c$ for 
all values of $n^*$
and all the minima in our study is well-represented by 
\begin{equation}
\Delta f_c(n^*,t) = \Delta f_0(n^*) + c(n^*) t^{-\alpha},
\label{fit}
\end{equation}
with $\alpha$ in the range $0.25-0.40$. 
Typical fits to this form with $\alpha = 0.35$
for two minima with $L$ = 15
and $L$ = 12 are shown in Fig.2. The values of $\Delta f_0$ obtained from such
fits with a fixed value of $\alpha$ are nearly independent of $n^*$, 
but exhibit a
dependence on the value of $\alpha$, varying between 0 and 0.5
for the $L$ = 15 minimum, and between 1.3 and 1.5 for the $L$ = 12 minimum 
of Fig.2. The quantity $c(n^*)$ increases with $n^*$. 
These results correspond to the function $g(n^*,x)$  having the form
$g(n^*,x) \sim A(n^*)x^{1/\alpha}$ with $A(n^*)$ 
decreasing with increasing $n^*$. 
We conclude from these observations that the 
growth of the effective barrier height
with increasing $n^*$ is primarily due to an 
entropic mechanism associated with an increase
of the difficulty in finding low-lying saddle points that connect different 
glassy local minimum of the free energy. This conclusion is consistent with the
canonical MC results of Ref.\cite{dv96} where we found that 
while the time 
scale of transitions between minima increases dramatically with $n^*$, 
the free energy increment at
the transition point remains essentially independent of $n^*$. 
This implies that 
the probability of finding a path to other minima for a 
fixed value of the free-energy
increment decreases as $n^*$ is increased.

Our results for the dependence of $\Delta f_c$ on $n^*$ 
are consistent with the Vogel-Fulcher
law\cite{vf} which assumes the following form \cite{wa81} for our system:
\begin{equation}
\Delta f_c(n^*) = a + b/(n^*_c - n^*),
\label{vfeq}
\end{equation}
where $a$ and $b$ are constants and $n^*_c$ is expected 
to be close to the random
close packing density $n^*_{rcp} \simeq 1.23$. There is some ambiguity 
in trying to
fit our data to this form because our values of $\Delta f_c$ 
depend weakly on the time $t$. However, the value of $n^*_c$ obtained
from fits of our data for $\Delta f_c(n^*, t)$ to Eq.(\ref{vfeq}) 
with fixed $a$ is
nearly independent of $t$. This is
consistent with the form of Eq.(\ref{fit}) if 
$a = \Delta f_0$ , $b \propto t^{-\alpha}$, and $c \propto 1/(n^*_c-n^*)$. 
$\Delta f_0$ is indeed nearly independent of
$n^*$, and
the $t$-dependence of $b$ and the $n^*$-dependence of $c$ 
are in agreement
with the other two conditions.  For the $L$ = 15 case,
we can fit the data for $\Delta f_c$ at $t$ = 15,000 to 
the form of Eq.(\ref{vfeq}) with
$a$ = 0 ($\Delta f_0=0)$. 
The best fit, shown in Fig.3, corresponds to $n^*_c = 1.225$,  very
close to the expected result. 
The best fit to the $L$ = 12 data with $a \simeq 1.0$ 
(the difference between the 
values of $\Delta f_0$ for the $L$ = 12 and $L$ = 15 minima is about 1.0)
also yields a similar value of $n^*_c$. 
So, we conclude that the observed growth of
the effective barrier height is consistent with the Vogel-Fulcher form. 
The increase in the effective barrier height as $n^*$ is increased from
0.94 to 1.06 is about 25$k_BT$, corresponding to a growth of the 
characteristic time
scale of about ten orders of magnitude. Thus, the range of time 
scales covered in
our study is comparable to that used in Vogel-Fulcher fits 
of experimental data, and
much wider than what can be achieved in standard MC or 
molecular dynamics simulations.

In summary, our study shows
that the density-dependence of the characteristic time scale for 
transitions between  
glassy local minima of the free energy of a dense hard sphere system
arises primarily from entropic effects associated
with the difficulty of finding low-lying paths that connect 
such minima.
The observed growth of the time scale is quantitatively
consistent with the Vogel-Fulcher form. 
To our knowledge, this is the first explicit demonstration of entropic 
effects leading
to a Vogel-Fulcher-type growth of relaxation times in a simple model liquid.
The same behavior should occur in other simple liquids where
$C(r)$ is similar\cite{hm86}. 

CD thanks the Supercomputer Education and Research Centre of
Indian Institute of Science for computational facilities and 
ICTP, Trieste, for hospitality.

\begin{figure}
\caption{The transition probability $P$ (see text)
as a function of the free-energy increment
$\Delta f$ for four values of the time $t$. 
The data shown are for a $L$ = 12 minimum
at $n^*$ = 1.04. 
The values of $\Delta f_c$ are indicated by the filled circles.} 
\label{fig1}
\end{figure}

\begin{figure}
\caption{Plots of $\Delta f_c$, obtained for a $L$ = 15 minimum, against 
$(t/1000)^{-0.35}$ for three values of $n^*$. The dashed lines are the
best straight-line fits.
Similar plots for a $L$ = 12 minimum are shown
in the inset.}
\label{fig2}
\end{figure}

\begin{figure}
\caption{The dependence of $\Delta f_c(n^*, t=15000)$ on $n^*$ for $L$ = 15. 
The dashed
line shows the best fit to a Vogel-Fulcher form (see text).}
\label{fig3}
\end{figure}


\begin{references}
\bibitem[*]{jnc} Also at the Condensed Matter Theory Unit, Jawaharlal Nehru Centre
for Advanced Scientific Research, Bangalore 560064, India.
\bibitem{rev} For a review, see {\it Liquids, Freezing and the Glass 
Transition}, edited by J.P. Hansen, D. Levesque, and J. Zinn-Justin,
(Elsevier, New York, 1991).
\bibitem{angell} C.A. Angell, J. Phys. Chem. Solids {\bf 49}, 863 (1988).
\bibitem{vf} H. Vogel, Z. Phys. {\bf 22}, 645 (1921); G.S. Fulcher, J.
Am. Ceram. Soc. {\bf 8}, 339 (1925).
\bibitem{kauz} W. Kauzmann, Chem. Rev. {\bf 48}, 219 (1948).
\bibitem{gdm} J.H. Gibbs and E.A. DiMarzio, J. Chem. Phys. {\bf 28}, 
373 (1958); G. Adams and J.H. Gibbs, J. Chem. Phys. {\bf 43}, 139 (1965).
\bibitem{ktw} T.R. Kirkpatrick and P.G. Wolynes, Phys. Rev. B {\bf 36}, 8552
(1987); T.R. Kirkpatrick, D. Thirumalai, and P.G. Wolynes, Phys.
Rev. A {\bf 40}, 1045 (1989).
\bibitem{ry} T.V. Ramakrishnan and M. Yussouff, Phys. Rev. B {\bf 19},
2275, (1979).
\bibitem{hm86} J.P. Hansen and I.R. McDonald, {\it Theory of Simple 
Liquids} (Academic, London, 1986).
\bibitem{cd92} C.Dasgupta, Europhys. Lett. {\bf 20}, 131 (1992).
\bibitem{lvd93} L.M. Lust, O.T. Valls, and C. Dasgupta, Phys. Rev. E {\bf 48},
1787, (1993).
\bibitem{dv94} C. Dasgupta and O.T. Valls, Phys. Rev. E {\bf 50}, 3916, (1994)
\bibitem{dv96} C. Dasgupta and O.T. Valls, Phys. Rev. E {\bf 53}, 2603,
(1996).
\bibitem{footn1} The term ``entropic'' refers to
the number of paths, which may be related\cite{ktw} to
the {\em configurational entropy}. It
is unrelated to the free energy being, in a different sense,
purely entropic in a hard sphere system.
\bibitem{wa81}L.V. Woodcock and C.A. Angell, Phys. Rev. Lett. {\bf 47},
1129 (1981).
\end{references}
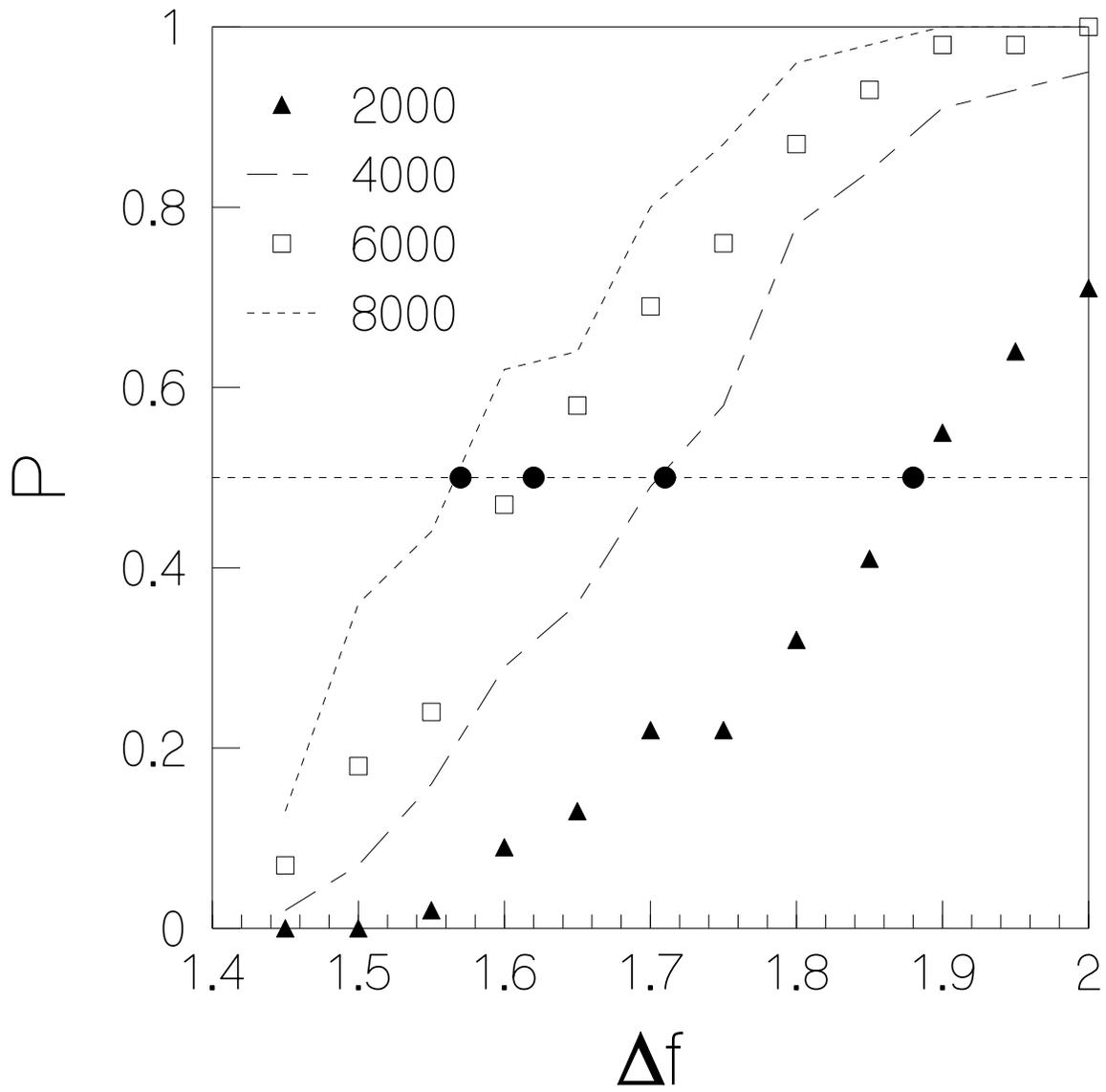
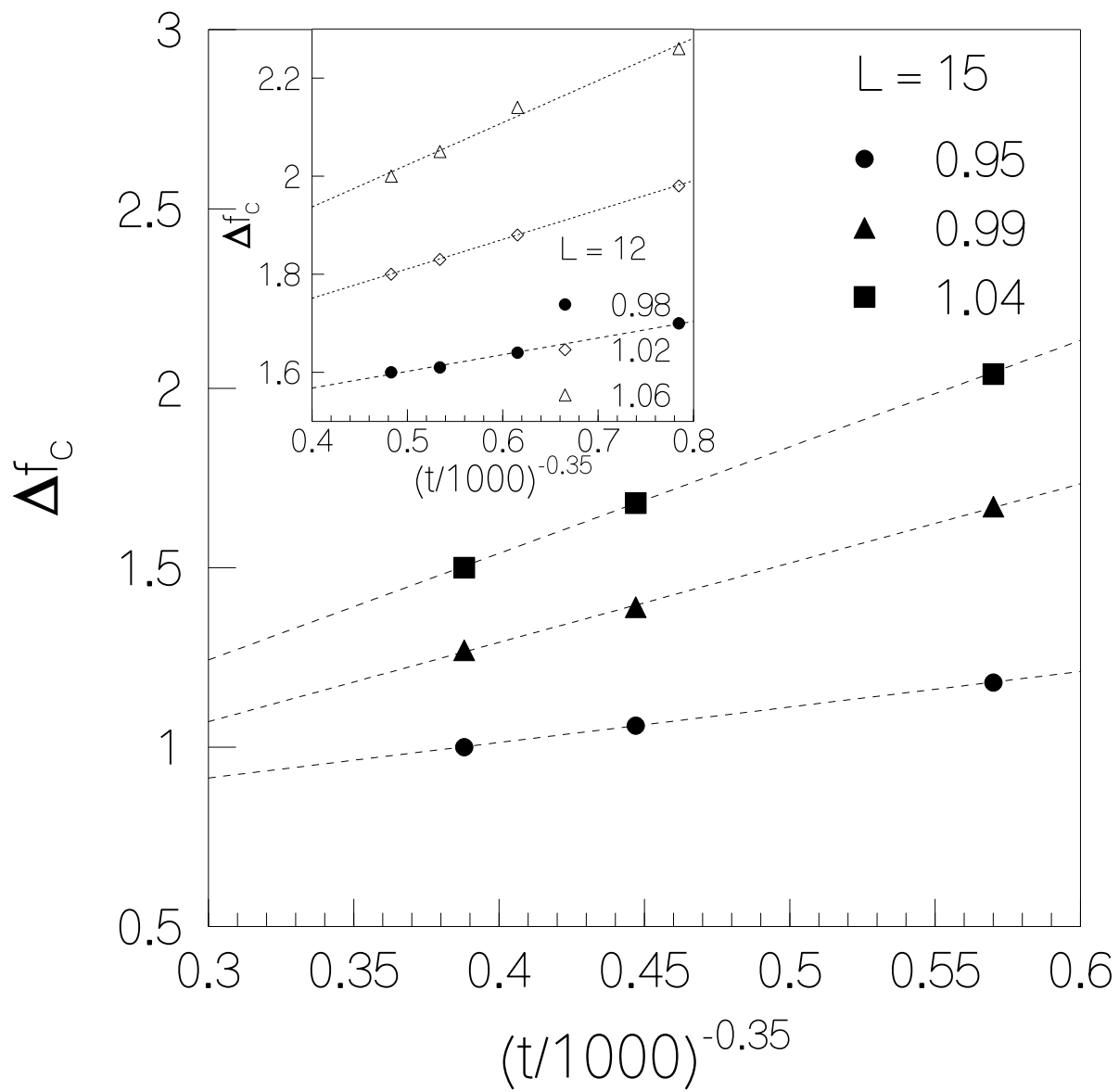
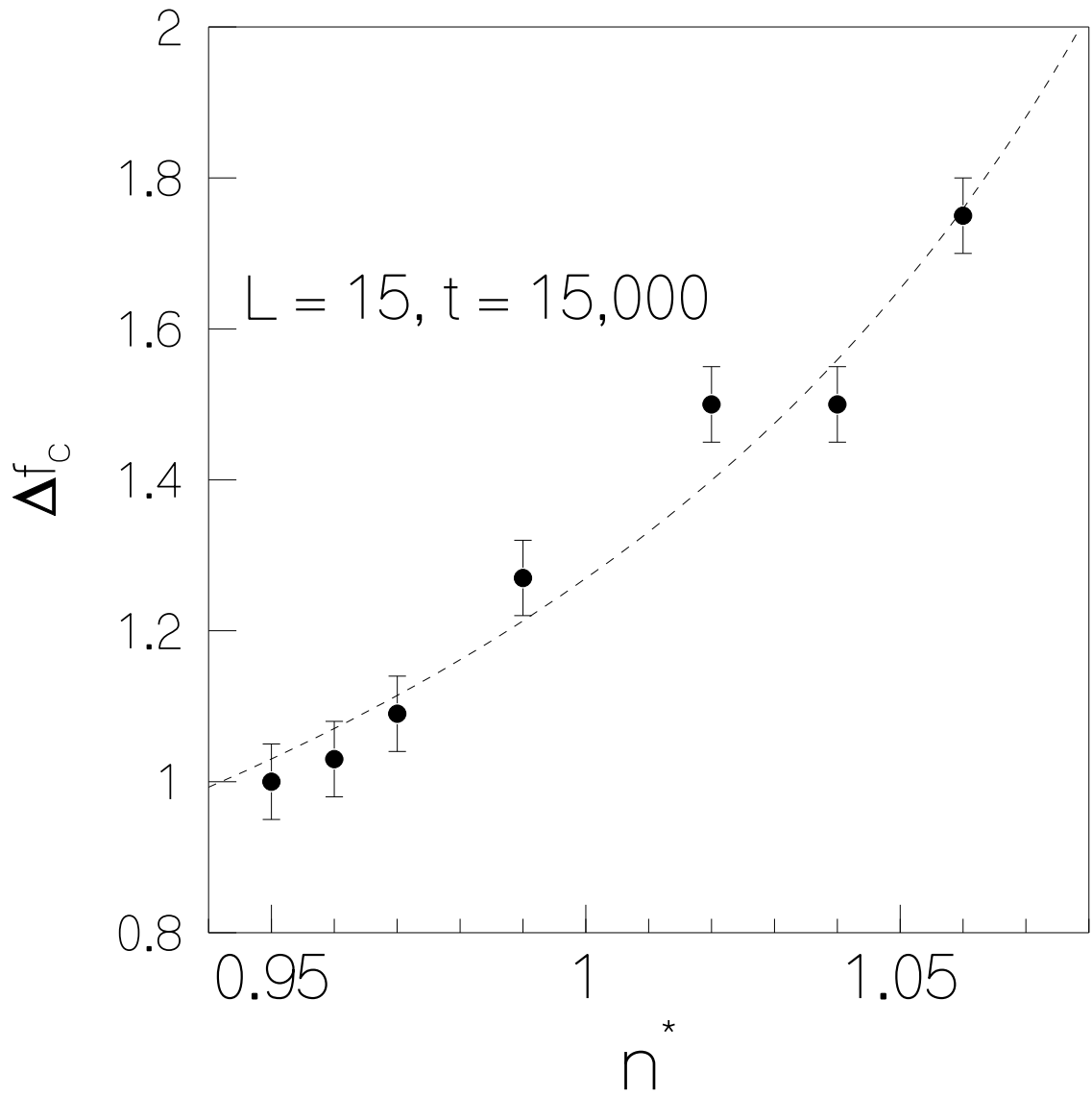
\end{document}